\documentclass[twocolumn,showpacs,amsmath,prl,amssymb,showkeys,superscriptaddress,aps,longbibliography,nofootinbib]{revtex4-2}

\pdfoutput=1

\usepackage{graphicx}
\usepackage{dcolumn}
\usepackage{mathtools}
\usepackage{scalerel}
\usepackage{bm}
\usepackage{mathrsfs}
\usepackage{amstext}
\usepackage{amsmath}
\usepackage{amssymb}
\usepackage[unicode=true,pdfusetitle,
 bookmarks=true,bookmarksnumbered=false,bookmarksopen=false,
 breaklinks=false,pdfborder={0 0 1},backref=false,colorlinks=true, citecolor=blue]
{hyperref}

\def\nuebar{\bar{\nu}_e} 

\def\nuAel{\nu {A}_{el}}

\def\DARpi{{\rm DAR} \mbox{-} \pi}
\def\q2{q^2}
\def\Enu{{\rm E}_{\nu}}
\def\xe135{^{135}{\rm Xe}}

\def\CRV{{\rm CR}^{\mbox{-}}}
\def\ACV{{\rm AC}^{\mbox{-}}}

\def\kgd{\rm kg\mbox{-}days}
\def\pcm2s1{\rm cm^{\mbox{-}2} s^{\mbox{-}1}}
\def\pkkd{\rm keV^{\mbox{-}1} kg^{\mbox{-}1} day^{\mbox{-}1}}

\def\VVB{{\rm \nuAel\mbox{-}candidate}}
\def\8ctg{{\rm AC^{\pm} {\otimes} CR^{\pm} {\otimes} B/S}}

\begin{document}

\title{
New Limits on Coherent Neutrino Nucleus Elastic Scattering Cross Section 
at the Kuo-Sheng Reactor Neutrino Laboratory
}

\newcommand{\as}{Institute of Physics, Academia Sinica,
Taipei 11529} 
\newcommand{\thu}{Department of Engineering Physics, Tsinghua University, 
Beijing 100084 }
\newcommand{\scu}{College of Physics, 
Sichuan University, Chengdu 610065}
\newcommand{\bhu}{Department of Physics, Institute of Science,
Banaras Hindu University, Varanasi 221005} 
\newcommand{\cusb}{Department of Physics,
School of Physical and Chemical Sciences,
Central University of South Bihar, Gaya 824236}  
\newcommand{\glau}{Department of Physics,
Institute of Applied Sciences and Humanities,
GLA University, Mathura 281406}  
\newcommand{ \hnbgu }{ Department of Physics, H.N.B. Garhwal University, 
Srinagar, Uttarakhand-246174} 
\newcommand{\deu}{Department of Physics,
Dokuz Eyl\"{u}l University, Buca, \.{I}zmir 35160} 
\newcommand{\itu}{Department of Physics Engineering,
Istanbul Technical University, Sarıyer, İstanbul 34467} 
\newcommand{\ndhu}{
Department of Physics, National Dong Hwa University,
Shoufeng, Hualien 97401} 

\newcommand{\corrhw}{htwong@phys.sinica.edu.tw}
\newcommand{\corrsk}{skarmakar@gate.sinica.edu.tw}
\newcommand{\corrms}{manu@gate.sinica.edu.tw}

\author{ S.~Karmakar }  \altaffiliation[Corresponding Author: ]{ \corrsk } \affiliation{ \as } \affiliation{ \glau }
\author{ M.K.~Singh }  \altaffiliation[Corresponding Author: ]{ \corrms } \affiliation{ \as } \affiliation{ \bhu }
\author{ V.~Sharma }  \affiliation{ \as } \affiliation{ \hnbgu }
\author{ H.T.~Wong } \altaffiliation[Corresponding Author: ]{ \corrhw } \affiliation{ \as }
\author{ Greeshma~C. }  \affiliation{ \as } \affiliation{ \cusb }
\author{ H.B.~Li }  \affiliation{ \as }
\author{ L.~Singh }  \affiliation{ \as } \affiliation{ \cusb }
\author{ M.~Agartioglu }  \affiliation{ \as } \affiliation{ \deu } \affiliation{ \ndhu }
\author{ J.H.~Chen } \affiliation{ \as }
\author{ C.I.~Chiang } \affiliation{ \as }
\author{ M.~Deniz } \affiliation{ \deu}
\author{ H.C.~Hsu }  \affiliation{ \as }
\author{ S.~Karada\v{g} } \affiliation{ \as } \affiliation{ \itu }
\author{ V.~Kumar }  \affiliation{ \as } \affiliation{ \glau }
\author{ C.H.~Leung }  \affiliation{ \as }
\author{ J.~Li }  \affiliation{ \thu }
\author{ F.K.~Lin } \affiliation{ \as }
\author{ S.T.~Lin } \affiliation{ \scu }
\author{ S.K.~Liu } \affiliation{ \scu }
\author{ H.~Ma }  \affiliation{ \thu }
\author{ K.~Saraswat } \affiliation{ \as }
\author{ M.K.~Singh } \affiliation{ \glau}
\author{ V.~Singh } \affiliation{ \cusb} \affiliation{ \bhu } 
\author{ D.~Tanabe } \affiliation{ \as }
\author{ J.S.~Wang } \affiliation{ \as }
\author{ L.T.~Yang } \affiliation{ \thu }
\author{ C.H.~Yeh }  \affiliation{ \as }
\author{ Q.~Yue } \affiliation{ \thu }

\collaboration{ TEXONO Collaboration }

\date{\today}

\begin{abstract}

Neutrino nucleus elastic scattering ($\nuAel$) 
with partial quantum-mechanical coherency
has been observed in several nuclei with
neutrinos from decay-at-rest pions.
This interaction with reactor neutrinos 
where coherency is $>$95\%
has not yet been experimentally observed.
We present new results on the studies of $\nuAel$ cross section
with an electro-cooled p-type point-contact germanium detector
at  the Kuo-Sheng Reactor Neutrino laboratory. 
A total of  
242(357)~$\kgd$ of Reactor ON(OFF) data
at a detector threshold of 200~eV in ionization energy are analyzed.
The Lindhard model parametrized by a single variable $k$ 
which characterizes the quenching function was used.
Limit at 90\% confidence level of $\rho {<} 4.7$ 
at the predicted value of $k {=} 0.162$
is derived, where $\rho$ is
the ratio of the measured to standard model (SM) cross section.
Conversely, $k {<} 0.288$ at the SM-value of $\rho {=} 1$. 

\end{abstract}


\maketitle


The  elastic scattering of neutrinos
with a nucleus $A(Z,N)$
(denoted by $\nuAel$, also represented as CE$\nu$NS in the 
literature)~\cite{DZFreedman:1974,nuA-komas}:
\begin{equation}
 \nu  ~ {+} ~  A(Z,N) ~  {\rightarrow} ~  \nu ~  {+} ~  A(Z,N)
\label{eq::nuAel}
\end{equation}
is a fundamental electroweak neutral current process
in the Standard Model (SM).
Its cross section at low momentum transfer ($\q2$)
is increased by quantum-mechanical coherency effects
relative to those of other neutrino interactions.
At typical reactor neutrino energy ($\Enu$) where
individual nucleon effects can be ignored,
the differential and total cross sections for $\nuAel$
can be approximated~\cite{Kerman:2016jqp} by, respectively:
\begin{eqnarray}
\label{eq::diffsigmanuA}
\left[ \frac{d \sigma } { d \q2 } \right]   & \approx  &
\frac{N^2}{2}  \left[ \frac{ G_F^2 }{ 4 \pi}  \right] 
\left[ 1 - \frac{ \q2 }{ 4 \Enu ^2 } \right] ~~~ {\rm and } \nonumber \\
\sigma & \approx & 
 N^2 \left[ \frac{ G_F^2  \Enu ^2}{ 4 \pi}  \right] 
\end{eqnarray}
showing an enhancement by $N^2$, the neutron number squared.
Studies of $\nuAel$~\cite{Abdullah:2022zue,Cadeddu_2023}
can provide sensitive probes to
physics beyond the SM (BSM), various astrophysical processes,
and the
neutron density distributions.
It can be applied to
supernova neutrino detection and
real-time monitoring of nuclear reactors
with compact and transportable neutrino detectors.
There are several active experimental programs on $\nuAel$,
with neutrinos from reactors or
from decay-at-rest pions ($\DARpi$) provided by
spallation neutron source~\cite{scholberg}.

First measurement of $\nuAel$
was achieved with $\DARpi$-$\nu$ in
the COHERENT experiment with
CsI(Na) scintillator~\cite{COHERENT_SCIENCE:2017},
followed by measurements with
liquid Ar~\cite{COHERENT:LAr:PRL2021} 
and Ge detectors~\cite{COHERENT:Ge:Adamski:2024yqt}.
The $\nuAel$ events from solar and atmospheric neutrinos are
the irreducible ``neutrino fog'' background
in dark matter experiments~\cite{PDG2024}.
First hints of observation of $\nuAel$ in
solar neutrinos are recently 
reported~\cite{PandaX-4T:PRL2024,XENONnT:PRL2024}.

The coherency effects of $\DARpi$-$\nu$, 
which can be quantitatively characterized by
   parameter $\alpha$~\cite{Kerman:2016jqp,TEXONO:2020vnv}, are only
   partial and incomplete ($\alpha {\sim} 14\% {-} 51\%)$.
These deviations from the perfect coherency case ($\alpha {=} 1$)
would have to be described and quantified before $\nuAel$
can be effectively applied towards other goals like sensitive BSM searches.
Reactor $\nuAel$, on the other hand, 
would typically have $\alpha {>} 95\%$ due to lower $\q2$. 
The nuclear recoil energy, however, is less than a few keV
which translates to the measurable ionization energy of $T {<} 500~{\rm eV}$
in germanium (Ge) detectors.
This poses severe constraints to the detector requirements.


Studies of reactor $\nuAel$ are actively pursued in several experiments.
Ge ionization detectors have the merits
of being a mature technology with relatively large modular mass, low detection threshold
and low intrinsic background. 
Ge-detectors with sub-keV sensitivities were first used in the 
TEXONO light Dark Matter~\cite{TEXONO:2007ccp} and $\nuAel$ 
program~\cite{Wong:2005vg}, 
which evolved to the adoption of the p-type Point-Contact 
Ge-detectors (pPCGe)~\cite{LUKE:1990-NIM-A,CoGeNT:PRL:2008,CoGeNT:PRL:2011,TEXONO:2016:AKSoma}.
Subsequent reactor $\nuAel$ experiments using pPCGe include 
CONUS~\cite{Bonet:2020awv,Bonet:2023kob,CONUS:2024lnu,PhysRevLett.133.251802},
DRESDEN-II~\cite{DRESDEN_PRL} and $\nu$GEN~\cite{NuGEN:PRD:2022}.
The pursuit of reactor $\nuAel$ with pPCGe catalyzed the CDEX Dark Matter program
at the China Jinping Underground Laboratory~\cite{Cheng:2017usi}.

The TEXONO research program~\cite{TEXONO:2018} conducts
experiments at the Kuo-Sheng Reactor Neutrino 
Laboratory (KSNL)~\cite{HT_Wong_2007PRD,MDeniz_PRD2010,TEXONO:2013hrh,LSingh_PRD2019}. 
The laboratory has an overburden of about 30~meter-water-equivalence
and is located at a distance of 28~m from Core No.1 of KSNL with a nominal thermal power
output of 2.9~GW. 
The shielding design is realized with 50~tons of materials including, 
from outside in,
plastic scintillator panels for cosmic-ray (CR) vetos,
15~cm of lead, 5~cm of stainless steel structure,
25~cm of boron-loaded polyethylene and 5~cm of oxygen 
free highly conducting copper.
Subject-specific experimental configurations are 
placed into an inner space with dimension of 
$60~{\rm cm} {\times} 100~{\rm cm} {\times} 75~{\rm cm (height)}$.
Details of the facility are documented 
in previous publications~\cite{HT_Wong_2007PRD,MDeniz_PRD2010}.

\begin{figure}
{\bf (a)}\\
\includegraphics[width=8.2cm]{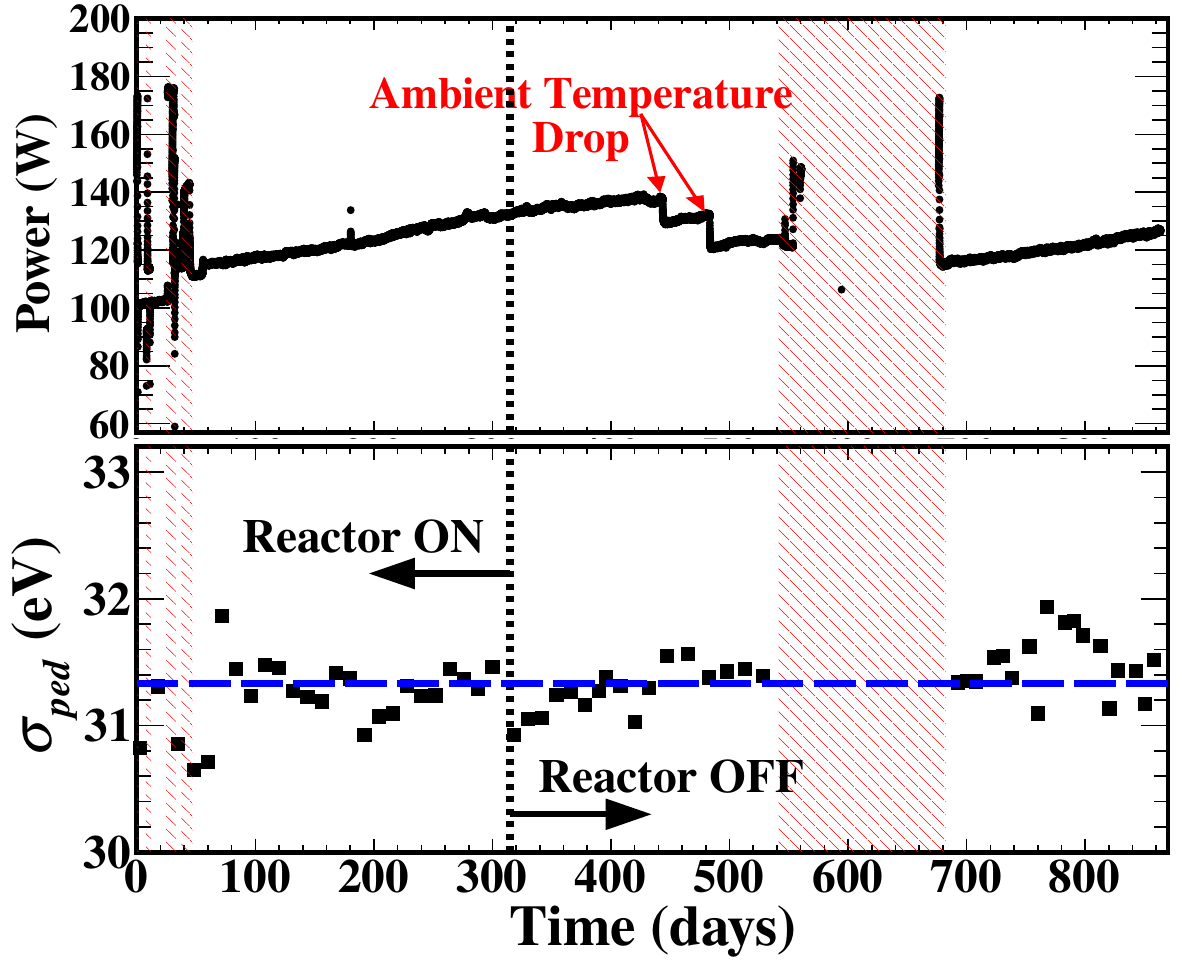}\\
{\bf (b)}\\
\includegraphics[width=8.2cm]{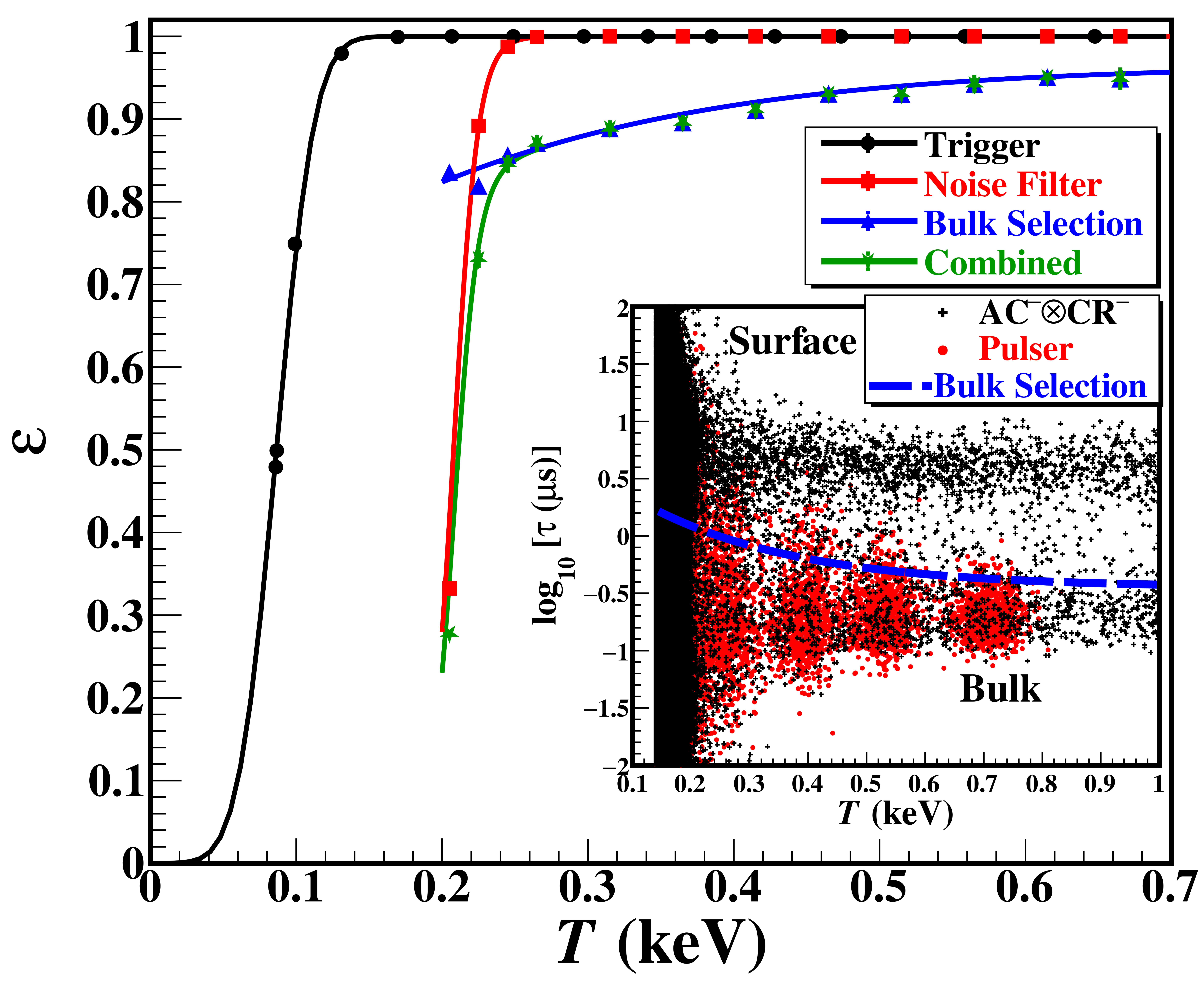}
\caption{
(a) Power consumption by the electro-cooler of pPCGe and the pedestal noise $\sigma_{ped}$
during data taking, showing $\sigma_{ped}$ is stable and
uncorrelated to the cooler operation.
Shaded bands are suspended periods of data taking.
All structures of the cryo-cooler power evolution are 
due to understood hardware or ambient conditions.
(b)
Measured signal efficiencies due to trigger conditions,
pedestal noise filters and B-Selection.
The $\tau$-distributions for physics and pulser events,
as well as the B(S) selection contour, are displayed in the inset.
The $T$-independent efficiencies are
listed in Table~\ref{tab::signaleff}.
}
\label{fig::dataquality}
\end{figure}

This analysis is based on data taken at KSNL 
from December 2020 to January 2023 with a pPCGe~\cite{pPCGe-Mirion}
enclosed by an Anti-Compton (AC) detector made of
NaI(Tl) scintillating crystal~\cite{HT_Wong_2007PRD}.
The target sensor is a Ge-crystal with diameter 70~mm and
height 70~mm and a mass of 1434~g. 
There are two novel features of this ``Generation-3 (G3)''
detectors relative to earlier ones: 
(i) electro-cooling of the 
Ge-sensors~\cite{CP5-CANBERRA,StirlingOP:2011} and 
(ii) improved front-end electronics design. 
In addition, ``pulser'' events are recorded for various control
and calibration purposes.
These are signals generated with 
programmable rise-times optimized to match those
of the intrinsic 10.37~keV cosmogenic peak due to $^{68}$Ge,
and with amplitudes adjusted to match the respective energy 
values~\cite{Wang:2024phr,Bonet:2023kob}.

Detector threshold for physics analysis is crucial
in the studies of reactor $\nuAel$. 
This is characterized by two measurables:
(i) the resolution in full-width-half-maximum (FWHM) with
pulser events, and
(ii) the ``noise-edge'' below which energy
the electronic pedestal noise would dominate the event rates.
The achieved pulser FWHM is 70.2~eV and the noise-edge 
is 200~eV for the G3 detectors.  
These improves over previous generations of pPCGe
with pulser FWHM at 150~eV and 122~eV
and noise-edge at 500~eV and 
300~eV~\cite{TEXONO:2013hrh,TEXONO:2016:AKSoma,LSingh_PRD2019,SINGH201963}.

\begin{figure}
{\bf (a)}\\
\includegraphics[width=8.2cm]{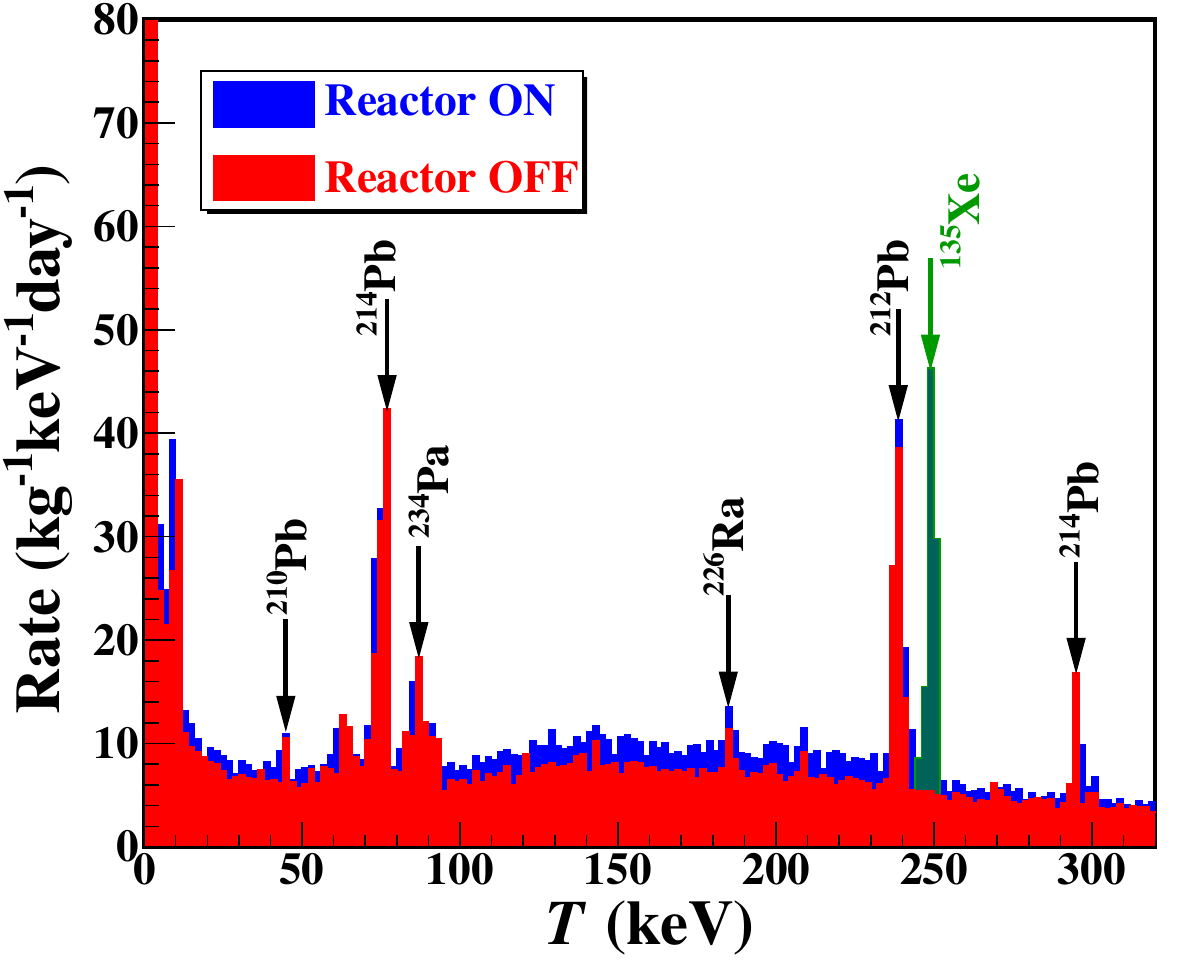}\\
{\bf (b)}\\
\includegraphics[width=8.2cm]{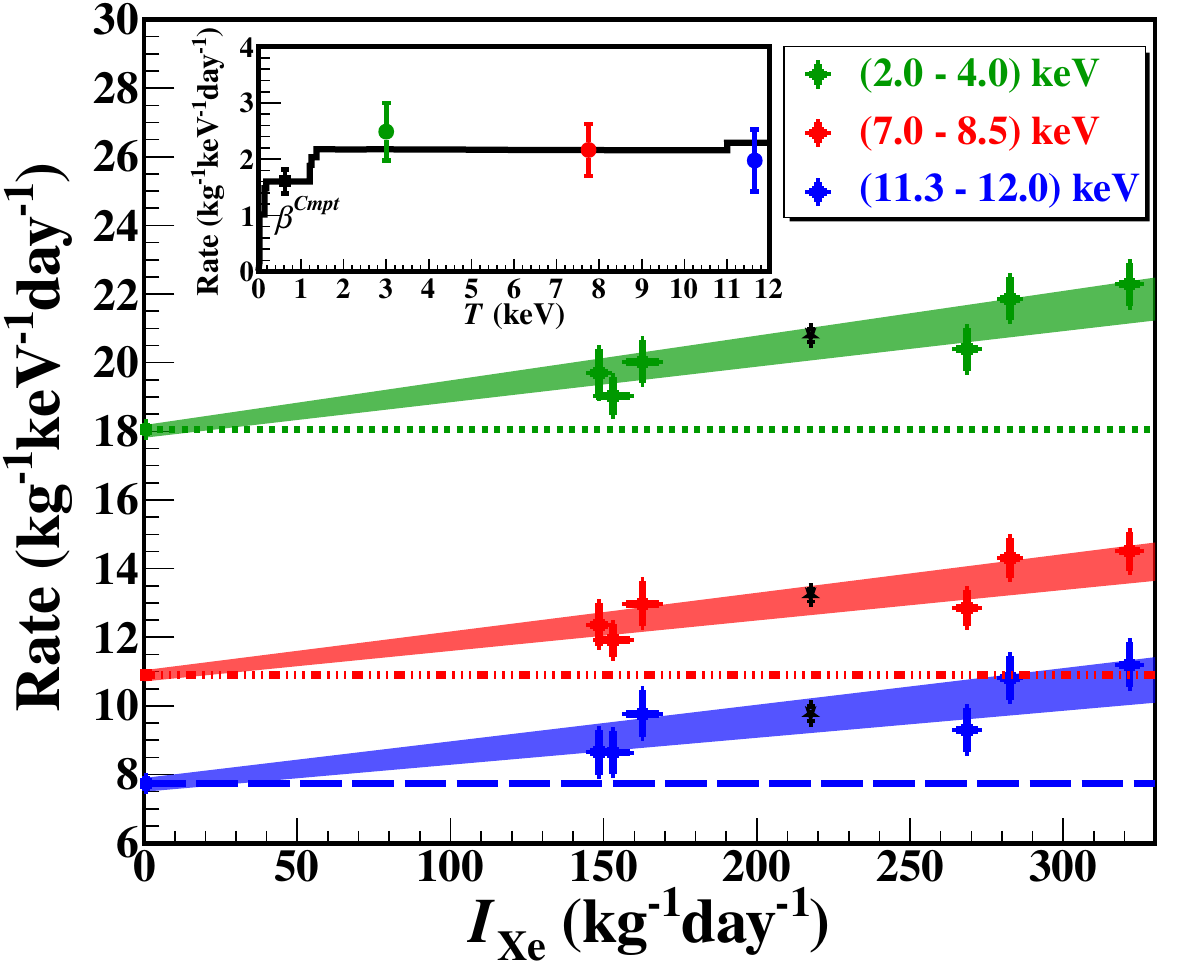}
\caption{
  (a) 
Measured high energy $\VVB$ 
  spectra of pPCGe during Reactor ON and OFF,
showing the ON-only 249.8~keV $\gamma$-line due to $^{135}$Xe.
  (b) 
The correlation between the constant background rates 
in three different energy bands without cosmogenic X-ray peaks 
with the intensity ($I_{\rm Xe}$) of the $\xe135$ $\gamma$-line. 
The intercepts at  $I_{\rm Xe} {=} 0$ 
correspond to the accurately-measured Reactor OFF rates.
The shaded region denotes the $\pm 1\sigma$ uncertainty bands
and the black points correspond to the measured values
of the combined data. 
The predicted Compton background~\cite{LinST:2020} 
fixed by the rates of the three bands is depicted in the inset.
}
\label{fig::Xenon}
\end{figure}

Stability of the root-mean-square pedestal noise ($\sigma_{ped}$)
behavior is crucial for this long duration
data taking, and is illustrated in Figure~\ref{fig::dataquality}a.
The power consumption for the electro-cooler is superimposed, 
showing that its inevitable variations 
have no effects on the pedestal stability. 
Following the noise-edge versus $\sigma_{ped}$ behavior depicted
in Figure~12a of Ref.~\cite{TEXONO:2016:AKSoma},
the residual fluctuations of ${<}$0.5~eV corresponds to
effective change of physics noise-edge of ${<}$3.5~eV, well within
one energy bin in analysis.

A data acquisition trigger is produced
via a discriminator threshold to the pPCGe pulse at $6~\mu {\rm s}$
shaping time. 
The trigger efficiency measured with pulser signals is displayed in
Figure~\ref{fig::dataquality}b.
After straightforward data quality filters, 
a total of (242)357~$\kgd$ Reactor ON(OFF) data are analyzed.
Valid events are placed into eight categories 
according to their CR, AC and B(S) signatures, 
where B(S) is the bulk-versus-surface events differentiation.
Signal selection by CR and AC vetos are applied, 
so that the survived $\VVB$ events are 
in anti-coincidence to the other detector components.
The B(S) events are identified by their fast(slow) $\tau$ 
via pulse shape discrimination 
techniques~\cite{TEXONO:2013bju,TEXONO:2016:AKSoma,Yang:2016crf}.
This is illustrated in the inset of Figure~\ref{fig::dataquality}b.
Signal efficiencies are derived by the survival of pulser events
under the same selection procedures.
The surface layer thickness is 
${\le} 0.5 ~ {\rm mm}$~\cite{pPCGe-Mirion}
and the fiducial mass is 1383~g.   

The non-$\nuAel$ background events are important: 
(i) to provide efficiency measurements on the various selection procedures,
(ii) as stability monitors, and
(iii) as tuning samples for parameter optimization in the analysis algorithms. 
The energy-independent signal detection efficiencies 
through the data acquisition and analysis procedures
are summarized in Table~\ref{tab::signaleff}.
The energy-dependent efficiencies from 
pedestal noise filters and BS selection
is depicted in Figure~\ref{fig::dataquality}b.
The analysis procedures follow closely the scheme established 
by previous pPCGe experiments described in 
Refs.~\cite{HT_Wong_2007PRD,TEXONO:2013bju,TEXONO:2016:AKSoma,Yang:2016crf}.


\begin{table}
 \centering
\caption{
Summary of $T$-independent
signal selection efficiencies.
The $T$-dependent components are displayed in 
Figure~\ref{fig::dataquality}b.
}
\begin{center}
\renewcommand{\arraystretch}{1.0}
\begin{tabular}{|l|c|}
\hline
Signal Selection                  & ~~ Efficiency (\%) \\ \hline
~~ Data Acquisition                & 99.77    \\ 
~~ Basic Data Quality             & 99.64     \\ 
~~ Cosmic-Ray Veto $\CRV$         & 88.57     \\ 
~~ Anti-Compton Veto $\ACV$       & 99.84      \\ \hline 
\multicolumn{1}{|r|}{Combined}    & 87.91    \\ \hline 
\end{tabular}
\end{center}
\label{tab::signaleff}
\end{table}



Anomalous Reactor ON related background due to $\xe135$ 
contamination is observed in this data set.
The isotope $\xe135$~\cite{Xe135} is a fission product and 
undergoes $\beta$-decay 
at a half-life of 9.14~hour to $^{135}{\rm Cs}^*$ 
which de-excites promptly by $\gamma$-rays emissions
to the long-lived $^{135}{\rm Cs}$ ground state. 
As displayed in Figure~\ref{fig::Xenon}a,
the signature $\gamma$-line at 249.8~keV 
is measured in the Reactor ON spectra, at a rate of 
$I_{\rm Xe} {=} ( 218 {\pm} 2 ) ~ {\rm kg^{\mbox{-}1}day^{\mbox{-}1}}$ 
from which observable features at low energy are produced.
This is much higher than that in our 
earlier experiment at KSNL~\cite{HT_Wong_2007PRD}, in which
$I_{\rm Xe} {=} ( 11.6 {\pm} 0.5 ) ~ {\rm kg^{\mbox{-}1}day^{\mbox{-}1}}$.
The main cause of the increase is 
the use of electro-cooling in pPCGe which replaces
purging of the inner volume with the boil-off of liquid nitrogen 
from conventional dewar.

The Compton continuum background 
is linearly correlated with $I_{\rm Xe}$. 
This is illustrated in Figure~\ref{fig::Xenon}b for the
2-4~keV, 7-8.5~keV, and 11.3-12~keV energy regions.
The averaged rates of the three bands (black data-points) show consistent 
excess over their respective Reactor OFF $I_{\rm Xe} {=} 0$ levels,
confirming that high energy ambient $\gamma$-rays  
produce Compton background
to the $\VVB$ samples in the low energy (${<}$10~keV) region.
The spectrum is flat but with step-wise
structures at atomic transition energies.
There are no other ON-only ambient background sources
besides $\xe135$, whose
event rate is only ${<}1.5\%$ of the Reactor ON background
at the low energy ${<}$300~eV region. 
This background can be accounted for and subtracted.

\begin{figure}
\includegraphics[width=8.2cm]{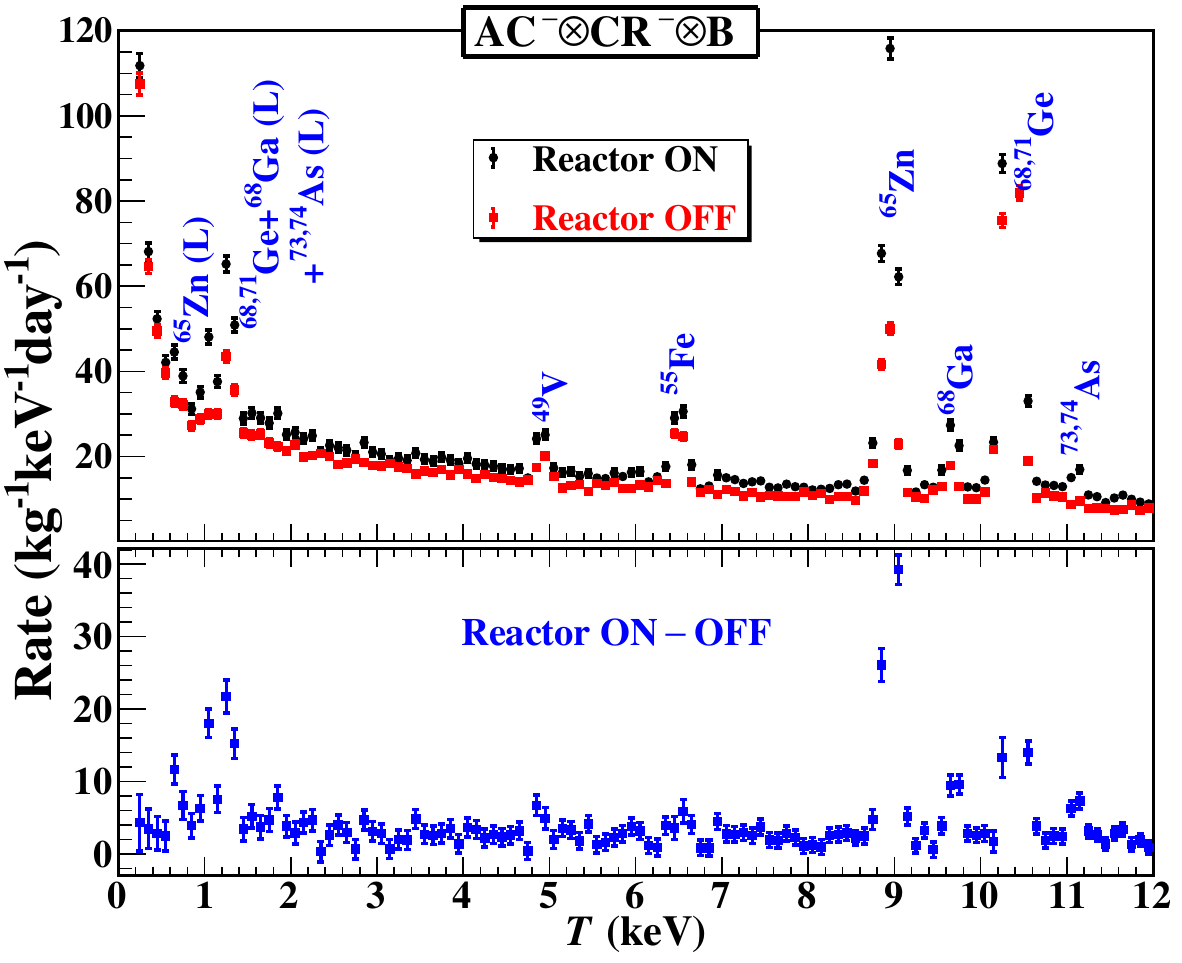}
\caption{
(Top) 
Candidate $\nuAel$ spectra from  
242(357)~$\kgd$ of Reactor ON(OFF) data. 
The cosmogenic X-ray lines are identified.
(Bottom) 
Reactor ON$-$OFF spectrum showing the 
finite $\xe135$ Compton excess. 
Cosmogenic peaks are observed for those
isotopes with half-lives comparable to the data taking
duration.
The 10.37~keV X-ray peak extends
off-scale and is truncated for clarity in display.
}
\label{fig::Residual_data}
\end{figure}

The reactor $\nuebar$-spectrum is adopted from that derived 
in Ref.~\cite{HT_Wong_2007PRD}. 
The total $\nuebar$-flux is fixed by the reactor thermal power output
and corresponds to $ 6.35 {\times} 10^{12} ~ \pcm2s1 $.
A conservative uncertainty of 5\% is assigned, 
at a comparable range with those of similar experiments~\cite{DRESDEN_PRL,PhysRevD.91.072001}.
As will be shown in what follows, 
while the neutrino flux is the leading systematic error in this analysis,
it is negligible compared to the measurement statistical uncertainties.
The Reactor ON/OFF and  ON$-$OFF spectra for
the $\VVB$ events at the low energy (${<}$12~keV) regions
are shown in Figure~\ref{fig::Residual_data}.
The constant excess in the ON$-$OFF spectrum is from
Compton background events due to $\gamma$'s
from $\xe135$. 
Its mean rate ($\beta^{Cmpt} {=} 1.61 ~ \pkkd$) 
and uncertainty ($\Delta^{Cmpt} {=} 0.21 ~ \pkkd$) 
at 200~eV are derived
using the Compton spectrum modeling of Ref.~\cite{LinST:2020}
and the best-fit ON$-$OFF measured background levels 
of the three energy ranges displayed in Figure~\ref{fig::Xenon}b.

\begin{figure}
{\bf (a)}\\
\includegraphics[width=8.2cm]{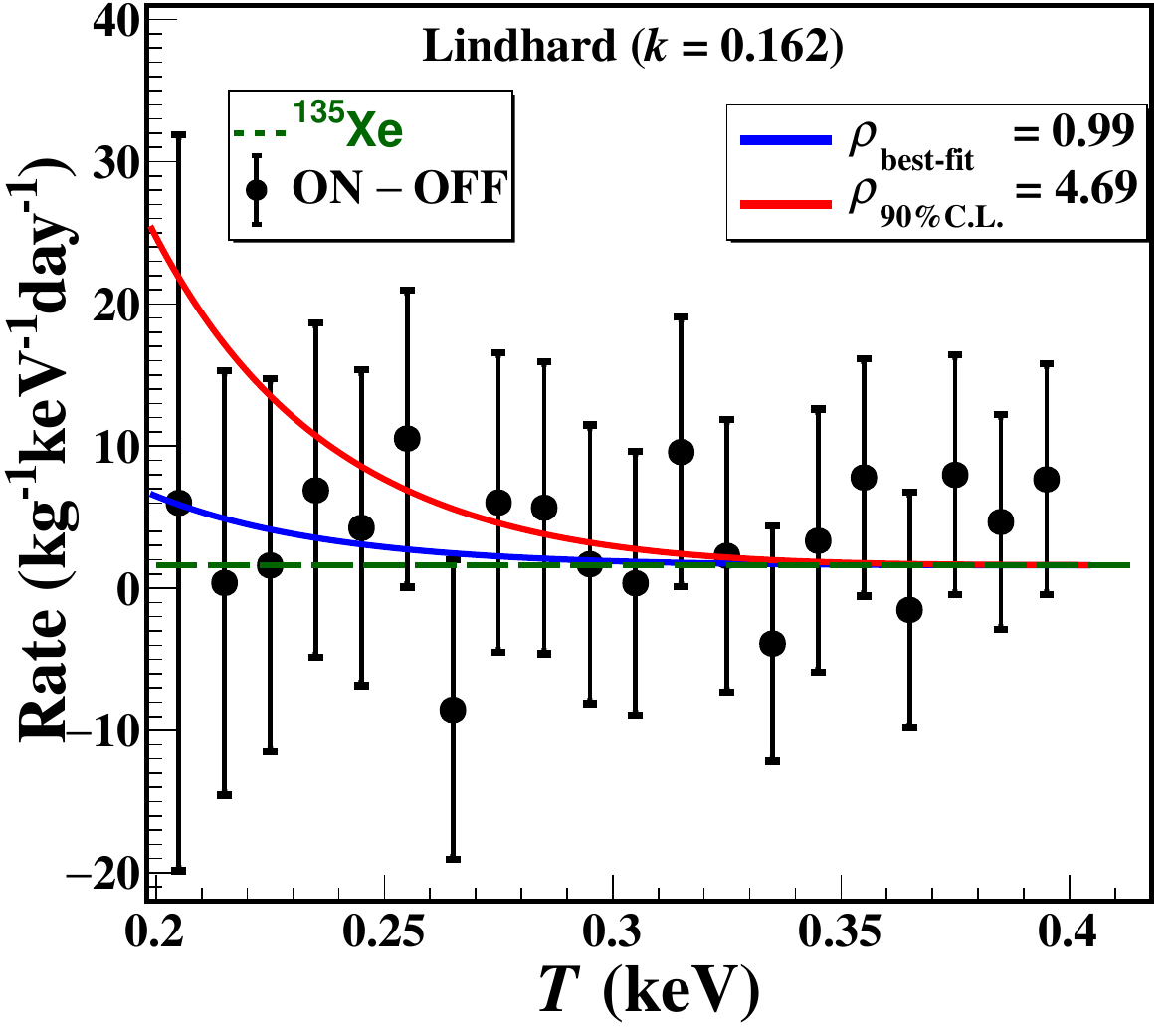}\\
{\bf (b)}\\
\includegraphics[width=8.2cm]{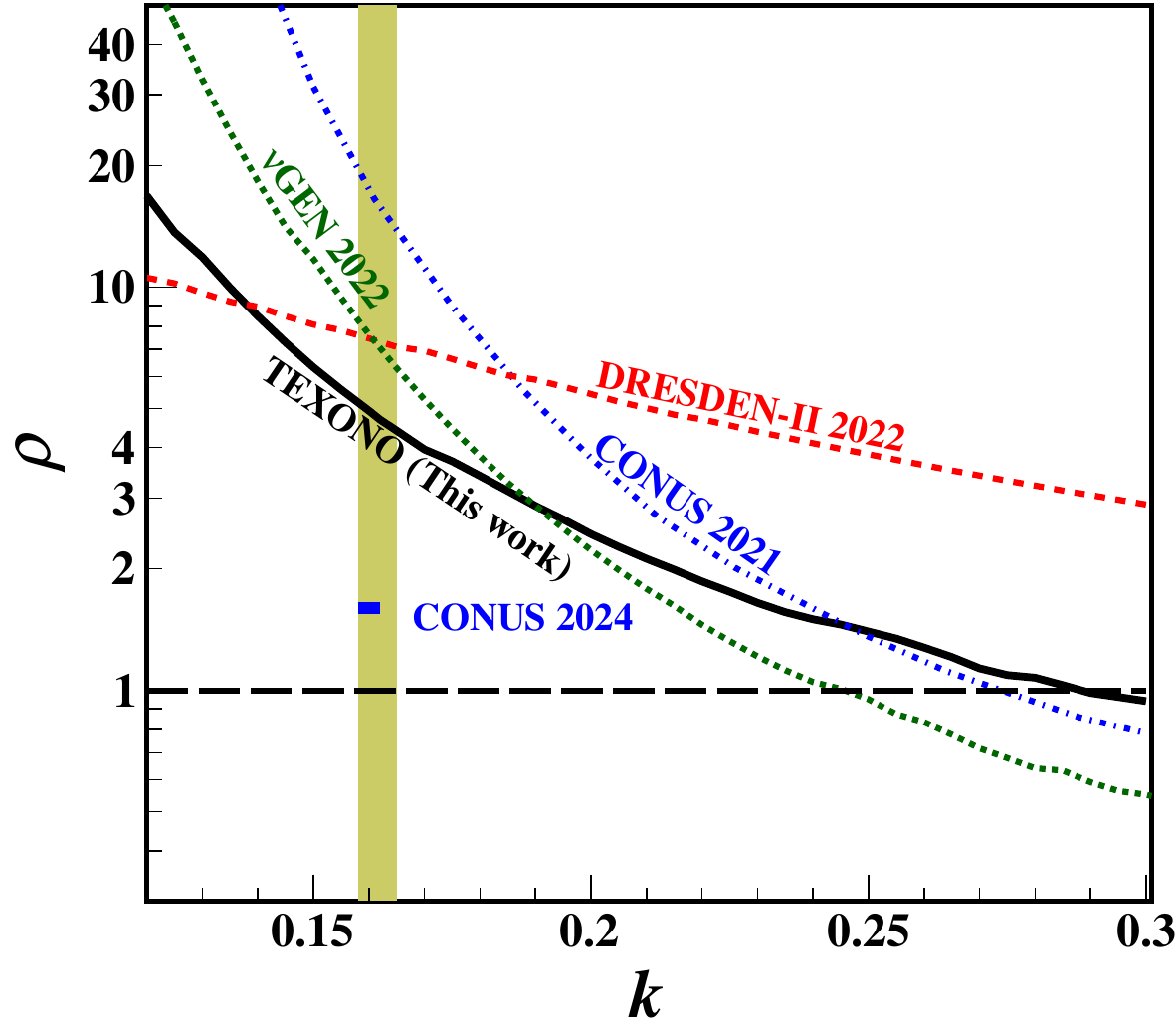}
\caption{
(a) 
The ON$-$OFF $\VVB$ spectrum at the relevant energy range.
Superimposed are the best-fit and 90\% C.L. upper limit 
spectra at $k {=} 0.162$ for Lindhard model.
The constant $\xe135$-induced background is depicted by the dashed line.
(b) 
Exclusion plot of $\rho$ versus $k$ in $\nuAel$ at 90\%~C.L. 
The $\pm 3 \sigma$ interval on $k$  
    is represented by the shaded band. 
Other reactor limits with pPCGe from 
CONUS~\cite{Bonet:2020awv,PhysRevLett.133.251802}, 
DRESDEN-II~\cite{DRESDEN_PRL} and
$\nu$GeN~\cite{NuGEN:PRD:2022}
are superimposed.
}
\label{fig::explot}
\end{figure}

Nuclear recoils are produced in $\nuAel$ 
and a ``quenching function'' is necessary to translate
this into the measurable energy deposition in Ge.
The benchmark Lindhard Model~\cite{Lindhard:1965} characterized by
a single parameter $k$ is adopted in this analysis.
The best estimate of $k {=} ( 0.162 {\pm} 0.003 )$ is derived from 
existing data~\cite{TEXONO:2016:AKSoma,Lindhard_data,Bonhomme:conus}. 
The anomalous low energy data 
of Ref.~\cite{DRESDEN-II-QF:PRD2021} 
with iron-filtered neutron are
excluded by the results of Ref.~\cite{PhysRevLett.133.251802}
while that with photo-neutron sources are in tension
with the recent measurements of Refs.~\cite{Bonhomme:conus,Liphd}.
They are therefore not adopted in this analysis.
A $\chi ^2$-analysis is applied 
to the the signal region of 200$-$400~eV at bin-size of 10~eV 
as a function of $k$: 
\begin{eqnarray}
  \chi^{2} ( \rho , \beta ; k) & = & 
\sum_{i}\left[ \frac{N_{i}-\rho ~ \nu^{\rm SM}_{i}(k)- \beta}
{\Delta_{i}}   \right]^{2}  \nonumber \\
& & ~~~~~ + \left[ \frac{\beta  -\beta^{Cmpt}}{\Delta^{Cmpt}} \right]^{2} ~~ ,
  \label{eqn:chi2}
\end{eqnarray}
where the $N_{i}$ ($\Delta_{i}$) are the counts (uncertainties) of
the $i^{th}$ data point of ON$-$OFF spectrum,
$\nu^{\rm SM}_{i}(k)$ is the SM event rate at Lindhard-$k$. 
The two outputs are: the best-fit estimate of the $\xe135$ level ($\beta$)
and the ratio of the experimental cross section relative
to that of SM $\nuAel$ ($\rho$). 
Depicted in Figure~\ref{fig::explot}a
are the ON$-$OFF spectrum with the best-fit results 
and the 90\% confidence level (C.L.) upper limit 
at $k {=} 0.162$ derived by the  
unified approach~\cite{Feldman_PRD98,JNeyman:1935}. 
The increased error bar at the threshold of 200~eV is
due to the drop in signal efficiency shown in Figure~\ref{fig::dataquality}b.
Figure~\ref{fig::explot}b depicts
the exclusion plot at 90\% C.L. in $\rho$ versus $k$, together with 
the constraints from the other reactor pPCGe experiments: 
the published CONUS results~\cite{Bonet:2020awv,PhysRevLett.133.251802},
as well as those from DRESDEN-II~\cite{DRESDEN_PRL} and
$\nu$GeN~\cite{NuGEN:PRD:2022} data derived via the same analysis procedures.

The best estimate of the $\xe135$ ON-only background at 200~eV is
$\beta {=} ( 1.62 {\pm} 0.22 )~ \pkkd$. 
The uncertainty is small relative to that of the measured spectrum
of $4.26~\pkkd$ within the interval of 200$-$280~eV
which contains 90\% of the $\nuAel$ signals in this analysis.
The effects of this background are therefore 
minor at the present level of sensitivity.
Improved sensitivities at $k {=} 0.162$ of 
$\rho {=} \left[ 0.99 {\pm} 2.23 (stat.) {\pm} 0.05 (sys.) \right]$ 
as the best-fitted value and
$\rho {<} 4.7$
as the 90\% C.L. upper limit were derived.
Alternatively, in the case where $\nuAel$ has SM cross section at
$\rho {=} 1.0$, then $k {<} 0.288$ at 90\% C.L.
We note also that
the CONNIE reactor $\nuAel$ experiment 
with CCD sensors (silicon) has provided a 
95\% C.L. upper limit of $\rho {<} 76$~\cite{connie2024}. 

We report in this Letter a new limit on reactor $\nuAel$ measurement at KSNL
which shows an improved sensitivity towards the goal of observation.
More data with comparable exposure were taken 
but under sub-optimal conditions during
the COVID-pandemic lockdown periods. 
Efforts are being made to recover these for physics analysis.
We continue to explore advanced pulse shape analysis software 
techniques with objectives
on reducing uncertainties in the bulk-surface event differentiation as well as 
pushing on the analysis threshold. 
The next generation ``G4'' pPCGe detectors
are expected to have the sensitivities of ${<}$50~eV in pulser FWHM 
and ${<}$150~eV in physics noise-edge, 
offering the exciting prospects of
positive $\nuAel$ observation at reactors.
All reactor cores of KSNL were decommissioned by March 2023. 
Our $\nuAel$ program will continue at the new Sanmen Reactor Laboratory 
under construction in Zhejiang, China~\cite{Yang:20249T}.\\
{\it Notes added after completion of this work: }
the CONUS+ experiment reports preliminary results of observing 
a 3.7$\sigma$ signature of the $\nuAel$ process 
at power reactor~\cite{Ackermann:2025obx}.

This work is supported by the Investigator Award AS-IA-106-M02 
and Thematic Project AS-TP-112-M01 from the Academia
Sinica, Taiwan, and  contracts 106-2923-M-001-006-MY5, 
107-2119-M-001-028-MY3 and 110-2112-M-001-029-MY3, 
113-2112-M-001-053-MY3
from the National Science and Technology Council, Taiwan.

\bibliography{nunkr-ref}

\end{document}